\documentstyle[12pt]{article}
\begin{document}
\tolerance=5000
\def\be{\begin{equation}}
\def\ee{\end{equation}}
\def\bea{\begin{eqnarray}}
\def\eea{\end{eqnarray}}
\def\nn{\nonumber \\}
\def\cF{{\cal F}}
\def\det{{\rm det\,}}
\def\Tr{{\rm Tr\,}}
\def\e{{\rm e}}
\def\etal{{\it et al.}}
\def\erp2{{\rm e}^{2\rho}}
\def\erm2{{\rm e}^{-2\rho}}
\def\er4{{\rm e}^{4\rho}}
\def\etal{{\it et al.}}

\  \hfill
\begin{minipage}{2.5cm}
NDA-FP-?? \\
\end{minipage}

\

\vfill

\begin{center}

{\large\bf Quantum (in)stability of 2D charged dilaton black holes 
and 3D rotating black holes}

\vfill

{\large\sc Shin'ichi NOJIRI}\footnote{
e-mail : nojiri@cc.nda.ac.jp}
and
{\large\sc Sergei D. ODINTSOV$^{\spadesuit}$}\footnote{
e-mail :
odintsov@tspi.tomsk.su}
\vfill

{\large\sl Department of Mathematics and Physics \\
National Defence Academy \\
Hashirimizu Yokosuka 239, JAPAN}

{\large\sl $\spadesuit$
Tomsk Pedagogical University \\
634041 Tomsk, RUSSIA \\
}

\vfill

{\bf ABSTRACT}

\end{center}

The quantum properties of charged black holes (BHs) in 2D dilaton-Maxwell 
gravity 
(spontaneously compactified from heterotic string)
 with $N$ dilaton coupled scalars are studied. 
We first investigate 2D BHs found by McGuigan, Nappi and Yost.  
Kaluza-Klein reduction of 3D gravity with minimal scalars 
leads also to 2D dilaton-Maxwell gravity with dilaton coupled scalars and 
the rotating BH solution found by Ba\~nados, Teitelboim and Zanelli (BTZ)
 which  
can be also described by 2D charged dilatonic BH. 
Evaluating the one-loop effective action for dilaton coupled scalars 
in large $N$ (and $s$-wave approximation for BTZ case), we show that 
quantum-corrected BHs may evaporate or else anti-evaporate similarly to 
4D Nariai BH as is observed by Bousso and Hawking. Higher modes may cause 
the disintegration of BH in accordance with recent observation by Bousso.

\ 

\noindent
PACS: 04.60.-m, 04.70.Dy, 11.25.-w

\newpage

\section{Introduction}

String theory opens new possibilities in the study of fundamental problems 
in black holes (BHs) physics, like string origin of BH entropy (for 
a review, see \cite{P}). Moreover, string duality dictates various 
relations between higher dimensional BHs and lower dimensional BHs,
as for example between 5D and 4D BHs (where D-branes methods have been 
applied to calculate the entropy \cite{HS}) from one side and 
3D and 2D BHs \cite{Hy,T} from another side. This fact indicates 
that applying results of (quantum) calculations in 2D  
one can learn about properties of higher dimensional BHs.

Another indication that such approach may be very useful comes from the 
study of back-reaction of quantized matter to 4D BH when spherical 
reduction is applied. Here, the resulting effective action turns out to 
be the effective action of 2D dilatonic gravity with dilaton coupled matter.

Recently, some interesting effects have been discovered in this 
direction. Working in large $N$ and $s$-wave approximation (quantum 
minimal scalars) Bousso and Hawking \cite{BH2} demonstrated that 
degenerate Schwarzchild-de Sitter (SdS) (Nariai) BH may not only
evaporate but also anti-evaporate. The effect of anti-evaporation 
has been confirmed in refs.\cite{NO2} for conformal matter where 
large $N$ approach and 4D anomaly induced effective action have been 
used. It has also been shown in refs.\cite{NO2} that some of 
anti-evaporating SdS BHs may be initially stable when Hartle-Hawking 
no boundary condition \cite{HH} is used. 

Further studying SdS BH perturbations due to quantum minimal scalars 
Bousso\cite{B} found new effect: proliferation of de Sitter space,
which means disintegration of this space into an infinite number of copies 
of itself (the instability of de Sitter space has been demonstrated 
in ref.\cite{PGP}).

It could be really interesting to understand if all these developments 
may be realized in other types of BHs. The natural candidate to think 
about is charged BH. Working in large $N$ approach for dilaton 
coupled quantum scalars we study here the quantum properties of 2D 
charged BHs. The corresponding classical solutions have been found in 
\cite{MNY} (see also ref.\cite{GP}). They may be considered as some 
compactifications of Type II string solutions. Moreover, very naturally 
they may be considered as 2D analog of Reissner-Nordstr\" om (RN) charged 
4D BH. Note that some properties of 2D charged BHs \cite{MNY} have been 
discussed in refs.\cite{NP,EO} (for the calculation of BH entropy in this 
case, see\cite{GP,NP}).

On the other hand, Kaluza-Klein reduction of 3D gravity with minimal 
scalars leads to 2D dilaton-Maxwell gravity with dilaton coupled scalars 
and the rotating BH solution found by Ba\~nados, Teitelboim and Zanelli 
(BTZ) can be also described by 2D charged dilatonic BH. 
BTZ black hole (BH)\cite{BTZ} attracts a lot of attention due to various 
reasons. In particulary, it is related via T-duality with a class 
of asymptotically flat black strings \cite{H} and via U-duality 
it is related \cite{Hy} with 4D and 5D stringy BHs \cite{HS} 
which are asymptotically flat ones. Hence, microscopically computing 
the entropy of BTZ BH \cite{SK} may be applied via duality relations 
for the computing of entropy for higher dimensional BHs. 
This fact is quite remarkable as above types of BHs look completely 
different from topological, dimensional or space-time points of view. 

BTZ BH being locally ${\rm AdS}^3$ without curvature singularity 
may be considered as a prototype for general CFT/AdS correspondence 
\cite{MW}. Indeed, 3D gravity has no local dynamics but BH horizon induces 
an effective boundary (actually, 2D WZWN model). Hence,quantum studies 
around BTZ BH may help in better understanding of above correspondence.

The paper is organised as follows. In the next section we consider 
2D dilaton-Maxwell gravity which represents toy model for 
4D or 5D Einstein-Maxwell theory and review its 2D charged BH 
solutions discussed by Mc-Guigan, Nappi and Yost. The one-loop 
effective action for dilaton coupled 2D scalar which is obtained by
spherical reduction from 4D or 5D minimal scalar is found. We 
work in large $N$ approximation. Quantum dynamics of 2D charged BHs is 
discussed in third section. In particulary, quantum corrected 2D 
BH is constructed. 
The analytical and numerical study of its perturbations is presented. 
It shows the existence of stable (evaporating) and unstable 
(anti-evaporating) modes. Calculation of quantum corrections to mass, 
charge, temperature and BH entropy is done in section 4. Due to duality 
with 5D stringy BH last result gives also quantum corrected BH 
entropy for 5D BH. After that we study quantum dynamics of BTZ BH 
due to 3D minimal scalars. We work in large $N$ and $s$-wave approach 
where minimal scalars are described as 2D dilaton coupled scalars. The 
quantum corrected version of BTZ BH is found. It is shown that it can 
also evaporate or anti-evaporate. Higher modes 
perturbations in the quantum spectrum for these two kinds of charged 
BHs are also briefly discussed.

\section{One-loop effective action}

We start from the action which has been considered by McGuigan, Nappi 
and Yost in ref.\cite{MNY}. As it has been shown in \cite{MNY}, this 
action follows from compactification of heterotic string theory:
\be
\label{cMNY}
S={1 \over 16\pi G}\int d^2x \sqrt{-g}\e^{-2\phi}
\left( R + 4 (\nabla \phi)^2 + 4\lambda^2 - 4 F_{\mu\nu}^2 \right)\ .
\ee
where $\lambda^2$ is cosmological constant, 
$F_{\mu\nu}=\partial_\mu A_\nu - \partial_\nu A_\mu$, $\phi$ is dilaton. 
Note that above action has the form typical for 4D or 5D Einstein-Maxwell 
theory spherically reduced to two dimensions. Hence, it can be considered 
as toy model to describe 4D or 5D BH with spherical reduction.

It is remarkable that action (\ref{cMNY}) has the classical solutions which 
correspond to the 2D charged black hole (2D analogue of 
Reissner-Nordstr\" om black hole) with multiple horizon.
In the explicit form its metric and dilaton look like \cite{MNY,GP}
\bea
\label{MNYmetric}
ds^2&=&-\left( 1-2m \e^{-2\lambda x} + q^2 \e^{-4\lambda x}\right)dt^2 \nn
&&+\left( 1-2m \e^{-2\lambda x} + q^2 \e^{-4\lambda x}\right)^{-1}dx^2 \nn
\e^{-2(\phi-\phi_0)}&=&\e^{2\lambda x}\ .
\eea
Here $q$ and $m$ are parameters related to the
charge and the mass of BH, respectively. 
The extremal solution is given by putting $q^2=m^2$. 
If we define new coordinates $r$ and $\tau$ by
\bea
\label{r}
\e^{-2\lambda x}&=&{m+ \epsilon \tanh 2\lambda r \over q^2} \nn
t&=&{q \over \epsilon}\tau \nn
\epsilon^2&\equiv&m^2 - q^2\ ,
\eea
$r\rightarrow +\infty$ corresponds to outer horizon and 
$r\rightarrow -\infty$ corresponds to inner horizon.
Taking the limit $\epsilon\rightarrow 0$, we obtain
\bea
\label{exmetric}
ds^2&=&{1 \over \cosh^2 \lambda r}\left(d\tau^2 - dr^2\right) \nn
\e^{-2(\phi-\phi_0)}&=&{1 \over q}\ .
\eea
Note that the dilaton field becomes a constant in the limit.

We will discuss now the quantum corrections induced by $N$ free conformally
invariant dilaton coupled scalars $f_i$ (no background scalars $f_i$):
\be
\label{sS}
S^f=-{1 \over 2}\int d^2x \sqrt{-g}\e^{-2\phi}
 \sum_{i=1}^N(\nabla f_i)^2\ .
\ee
Note that the above action (\ref{sS}) appears as a result of
spherical reduction from 4D or 5D minimal scalar.

We present now the two-dimensional metric as the following:
\be
\label{2dmetric}
ds^2=\e^{2\sigma}\tilde g_{\mu\nu}d x^\mu d x^\nu
\ee
where in conformal gauge $\tilde g_{\mu\nu}$ is the flat metric.
Then the total quantum action coming from quantum scalars 
is given as (we work in large $N$ approximation what 
justifies the neglecting of proper quantum gravitational corrections):
\be
\label{Gamma}
\Gamma = W + \Gamma[1, \tilde g_{\mu\nu}]
\ee
where $W$ is conformal anomaly induced effective action \cite{BH,NO} 
and $\Gamma[1, \tilde g_{\mu\nu}]$ is one-loop effective action for scalars 
(\ref{sS}) calculated on the metric (\ref{2dmetric}) with $\sigma=0$ (the 
general covariance should be restored after all). Note that general 
structure of conformal anomaly has been discussed in ref.\cite{DS} while 
quantum conformal field theory in general dimensions (bigger than two) has 
been discussed in ref.\cite{EHO} (for earlier work,see \cite{JLC}). It 
could be interesting to generalize above works for the case when external 
dilaton presents.

Following \cite{NO} (due to corresponding result for conformal anomaly of 
(\ref{sS})\cite{NO,BH,A}), we get
\be
\label{qc}
W=-{1 \over 2}\int d^2x \sqrt{-g} \left[ 
{N \over 48\pi}R{1 \over \Delta}R  -{N \over 4\pi} \nabla^\lambda \phi
\nabla_\lambda \phi {1 \over \Delta}R +{N \over 4\pi}\phi R \right]\ .
\ee
We are left with the calculation of the conformally invariant part of the 
effective action $\Gamma[1, \tilde g_{\mu\nu}]$. This term has to be 
calculated on the flat space, after that the general covariance should be 
restored. It is impossible to do such calculation in closed form. One has 
to apply some kind of expansion. Using Schwinger-De Witt type expansion, we 
get this effective action as local curvature expansion.

Using the results of ref.\cite{NO}, one can obtain:
\be
\label{Gamma1}
\Gamma[1, \tilde g_{\mu\nu}]={N \over 24\pi}\int d^2x 
\sqrt{-\tilde g}(-6\tilde \nabla_\mu \phi 
\tilde \nabla^\mu \phi ) \ln \mu^2 + \cdots 
\ee
here $\mu$ is dimensional parameter.\footnote{Note that more exactly, one 
should write $\ln {L^2 \over \mu^2}$ where $L^2$ is covariant cut-off 
parameter.} Now, one has to generalize $\Gamma[1, \tilde g_{\mu\nu}]$ 
in order to present it in general covariant form:
\be
\label{Gamma1b}
\Gamma[1, g_{\mu\nu}]=-{N \over 4\pi}\int d^2x  
\sqrt{- g}\nabla_\mu \phi \nabla^\mu \phi \ln \mu^2 + \cdots \ .
\ee
This action is conformally invariant as it should be in accordance with 
Eq.(\ref{Gamma}) where $W$ gives scale-dependent part of total one-loop 
effective action. Note that terms which are not written explicitly in 
Eq.(\ref{Gamma1b}) are conformally invariant, higher derivatives terms on 
$\phi$. Moreover, as it is easy to see for minimal scalars $\phi=0$, 
$\Gamma[1, \tilde g_{\mu\nu}]=0$. That corresponds to general result that 
in two dimensions, effective action (\ref{Gamma}) is defined completely by 
only anomaly induced action $W$ (which is not the case in the presence of 
dilaton).

It is also interesting to note that if one uses some form of non-local 
expansion in the calculation of $\Gamma[1, \tilde g_{\mu\nu}]$, one would 
get the leading terms of this expansion similar to some terms in 
Eq.(\ref{qc}). In other words, such procedure would just decrease some of 
coefficientes in Eq.(\ref{qc}) (apart from appearence of few local terms.)

\section{Quantum dynamics of 2D charged BHs}

Let us work in the conformal gauge
\be
\label{cg}
g_{\pm\mp}=-{1 \over 2}\e^{2\rho}\ ,\ \ g_{\pm\pm}=0
\ee
The equations of motion with account of quantum corrections are given by 
variations of sum $S$(\ref{cMNY})$+W$(\ref{qc})$+\Gamma$(\ref{Gamma1b}) 
over $g^{\pm\pm}$, $\rho$, $\phi$ and $A_\mu$:
\bea
\label{eqnpp}
0&=&{1 \over 8G}\e^{-2\phi}\left(4\partial_\pm \rho \partial_\pm\phi 
- 2 \partial_\pm^2\phi\right) +{N \over 12}\left( \partial_\pm^2 \rho 
- \partial_\pm\rho \partial_\pm\rho \right) \nn
&& +{N \over 2} \left\{\left( \partial_\pm \phi 
\partial_\pm\phi \right)\rho+{1 \over 2}{\partial_\pm \over \partial_\mp}
\left( \partial_\pm\phi \partial_\mp\phi \right)\right\} \nn
&& +{N \over 4}\left\{ 2 \partial_\pm \rho \partial_\pm \phi 
-\partial_\pm^2 \phi \right\} + t^\pm(x^\pm) \nn
&& +{N \over 2}\left(\partial_\pm \phi\partial_\pm\phi \right)\ln \mu^2 \\
\label{req}
0&=&{1 \over 8G}\e^{-2\phi}\left(2\partial_+
\partial_- \phi -4 \partial_+\phi\partial_-\phi 
- \lambda^2 \e^{2\rho} + \e^{-2\rho}(F_{+-})^2\right) \nn
&& -{N \over 12}\partial_+\partial_- \rho
-{N \over 4}\partial_+ \phi \partial_- \phi 
+{N \over 4}\partial_+\partial_-\phi \\
\label{eqtp}
0&=& {1 \over 8G}\e^{-2\phi}\left(-4\partial_+
\partial_- \phi +4 \partial_+\phi\partial_-\phi +2\partial_+\partial_- \rho
+ \lambda^2 \e^{2\rho}+ \e^{-2\rho}(F_{+-})^2\right)  \nn
&& +2N\left\{-{1 \over 8}\partial_+(\rho \partial_-\phi)
-{1 \over 8}\partial_-(\rho \partial_+\phi)
-{1 \over 8}\partial_+\partial_-\rho \right\} \nn
&& -{N \over 2} \partial_+ \partial_- \phi \ln \mu^2\\
\label{eqA}
0&=&\partial_\pm(\e^{-2\phi-2\rho}F_{+-})\ .
\eea
Eq.(\ref{eqA}) can be integrated to give
\be
\label{eqA2}
\e^{-2\phi-2\rho}F_{+-}=B\ (\mbox{constant})\ .
\ee
We first consider the static case and replace $\partial_\pm$ by $\pm {1 
\over 2}\partial_r$. Then, using (\ref{eqA2}), Eqs.(\ref{eqnpp}), 
(\ref{req}) and (\ref{eqtp}) are rewritten as follows:
\bea
\label{eqnpp2}
0&=&{1 \over 8G}\e^{-2\phi}\left(\partial_r \rho \partial_r \phi 
-{1 \over 2} \partial_r^2\phi\right) +{N \over 48}\left( \partial_r^2 \rho 
- \partial_r\rho \partial_r\rho \right) \nn
&& +{N \over 8} \left( \rho + {1 \over 2}\right)
\partial_r \phi \partial_r\phi 
+{N \over 16}\left\{ 2 \partial_r \rho \partial_r \phi 
-\partial_r^2 \phi \right\} + t_0 \nn
&& +{N \over 8} \partial_r \phi \partial_r \phi \ln \mu^2\\
\label{req2}
0&=&{1 \over 8G}\e^{-2\phi}\left(-{1 \over 2}\partial_r^2
\phi + \partial_r\phi\partial_r\phi 
- \lambda^2 \e^{2\rho} + B^2\e^{2\rho+4\phi}\right) \nn
&& +{N \over 48}\partial_r^2 \rho
+{N \over 16}\partial_r \phi \partial_r \phi 
-{N \over 16}\partial_r^2\phi \\
\label{eqtp2}
0&=& {1 \over 8G}\e^{-2\phi}\left(\partial_r^2\phi 
- \partial_r\phi\partial_r\phi -{1 \over 2}\partial_r^2 \rho
+ \lambda^2 \e^{2\rho}+ B^2\e^{2\rho;4\phi}\right)  \nn
&& +{N \over 8}\partial_r(\rho \partial_r\phi)
+{N \over 16}\partial_r^2\rho + {N \over 8}\partial_r^2 \phi \ln \mu^2 \ .
\eea
Since the classical extremal solution can be characterized by constant 
dilaton field $\phi$, we assume $\phi$ is constant even when we include 
the quantum correction:
\be
\label{constphi}
\phi=\phi_0\ (\mbox{constant})\ .
\ee
Then (\ref{req2}) and (\ref{eqtp}) have the following forms:
\bea
\label{req3}
0&=&\e^{-2\phi_0}\left(-\lambda^2 + B^2\e^{4\phi_0}\right)
\e^{2\rho}+{GN \over 6}\partial_r^2\rho \\
\label{eqtp3}
0&=&\e^{-2\phi_0}\left(\lambda^2 + B^2\e^{4\phi_0}\right)
\e^{2\rho}+\left(-{1 \over 2}\e^{-2\phi_0}
+{GN \over 2}\right)\partial_r^2\rho \ .
\eea
The condition that Eqs.(\ref{req3}) and (\ref{eqtp3}) are compatible with 
each other gives
\be
\label{A}
B^2={\lambda^2\e^{-4\phi_0}\left(\e^{-2\phi_0}-{4GN \over 3} 
\right) \over \e^{-2\phi_0}-{2GN \over 3} } 
\ee
and we find the scalar curvature $R$ is constant
\be
\label{constR}
R=-2\e^{2\rho}\partial_r^2\rho=R_0
\equiv -{8\lambda^2 \e^{-2\phi_0} \over 
\e^{-2\phi_0}-{2GN \over 3} }\ .
\ee
We see that quantum corrections decrease the value of curvature if compare 
with the classical case. (\ref{constR}) can be integrated to give
\be
\label{rho0}
\e^{2\rho}=\e^{2\rho_0}\equiv {2C \over R_0}
\cdot {1 \over \cosh^2 \left(r\sqrt{C} \right)}\ .
\ee
Here $C$ is a constant of the integration. Finally (\ref{eqnpp2}) gives
\be
\label{t_0}
t_0={GN \over 6}C\ .
\ee

We now investigate the (in)stability of the above obtained extremal 
solution by the perturbation
\be
\label{pert}
\rho=\rho_0 + \epsilon R\ ,\ \ \ \phi=\phi_0 +\epsilon S\ .
\ee
Here $\epsilon$ is an infinitesimally small constant. By following Bousso 
and Hawking's argument \cite{BH2}, we neglect the second term in 
(\ref{qc}). (Note that we keep similar term with $ln \mu^2$ as it could be 
large). Then (\ref{req}) and (\ref{eqtp}) (and (\ref{eqA2})) give 
the following equations:
\bea
\label{reqpert}
0&=&{1 \over 8G}\e^{-2\phi_0}\left\{2\partial_+\partial_-S 
- 2\lambda^2\e^{2\rho_0} (R-S) + 2B^2 \e^{2\rho_0+4\phi_0}(R+S)\right\} \nn
&& -{N \over 12}\partial_+\partial_- R+ {N \over 4}\partial_+\partial_-S \\
\label{eqtppert}
0&=&{1 \over 8G}\e^{-2\phi_0}\left\{-4\partial_+\partial_-S 
+ 2\partial_+\partial_-R -4\partial_+\partial_-\rho_0 S
+ 2\lambda^2 \e^{2\rho_0} (R-S)\right.\nn
&& \left. + 2B^2 \e^{2\rho_0+4\phi_0}
(R+S)\right\} - {N \over 4}\partial_+\partial_- R 
-{N \over 2}\ln \mu^2 \partial_+\partial_- S \ .
\eea
We are going to study (in)stability of black holes in the same way as in 
four dimensions \cite{NO2}. As in the previous works, we assume $R$ and 
$S$ have the following form
\bea
\label{RS}
R(t,r)&=&P\cosh t\alpha\sqrt{C}\cosh^\alpha r\sqrt{C} \nn
S(t,r)&=&Q\cosh t\alpha\sqrt{C}\cosh^\alpha r\sqrt{C} \ .
\eea
Then we obtain the following equations
\bea
\label{eigenRS}
\cosh^2\left(r\sqrt{C}\right)\partial_+\partial_- R &=& A R \nn
\cosh^2\left(r\sqrt{C}\right)\partial_+\partial_- S &=& A S \nn
A&\equiv& {\alpha (\alpha - 1) C \over 4}\ .
\eea
Note that there is one to one correspondence between $A$ and $\alpha$ if 
we restrict $A>0$ and $\alpha<0$. Then (\ref{reqpert}) and 
(\ref{eqtppert}) become algebraic equations:
\bea
\label{alg1}
0&=&\left\{\e^{-2\phi_0}\left({R_0 \over C}A + 2\lambda^2
+ 2 B^2\e^{4\phi_0}\right) + {GNR_0 \over C}A\right\} Q \nn
&& + \left\{\e^{-2\phi_0}\left(- 2\lambda^2
+ 2 B^2\e^{4\phi_0}\right) - {GNR_0 \over 3C}A\right\} P \\
\label{alg2}
0&=&\left\{\e^{-2\phi_0}\left(-{2R_0 \over C}A -{R_0 \over 2}
- 2\lambda^2 + 2 B^2\e^{4\phi_0}\right)
-2GN (\ln\mu^2) {R_0 \over C}A\right\} Q \nn
&& + \left\{\e^{-2\phi_0}\left({R_0 \over C}A + 2\lambda^2
+ 2 B^2\e^{4\phi_0}\right) - {GR_0 \over C}A\right\}P\ .
\eea
In order that Eqs.(\ref{alg1}) and (\ref{alg2}) have a non-trivial solution 
for $Q$ and $P$, we find
\bea
\label{det}
F(A)&\equiv&\left\{\e^{-2\phi_0}\left({R_0 \over C}A + 2\lambda^2
+ 2 B^2\e^{4\phi_0}\right) + {GR_0 \over C}A\right\} \nn
&& \times \left\{\e^{-2\phi_0}\left({R_0 \over C}A + 2\lambda^2
+ 2 B^2\e^{4\phi_0}\right) - {GNR_0 \over C}A\right\} \nn
&& - \left\{\e^{-2\phi_0}\left(- 2\lambda^2
+ 2 B^2\e^{4\phi_0}\right) - {GNR_0 \over 3C}A\right\} \nn
&& \times \left\{\e^{-2\phi_0}\left(-{2R_0 \over C}A -{R_0 \over 2}
- 2\lambda^2 + 2 B^2\e^{4\phi_0}\right) \right. \nn
&& \left. -2GN (\ln\mu^2) {R_0 \over C}A\right\}=0\ .
\eea
Eq. (\ref{det}) should be solved with respect to $A$. 

When we compare the model with the 4D one, $\e^{-\phi}$ is identified with 
the radius coordinate. Since the (apparent) horizon is a null surface, 
the horizon is given by the condition
\be
\label{horizon}
\nabla\phi\cdot\nabla\phi=0\ .
\ee
Substituting (\ref{RS}) into (\ref{horizon}), we find the horizon is given 
by
\be
\label{horizonb}
r=\pm\alpha t\ .
\ee
Therefore on the horizon, we obtain
\be
\label{Shrzn}
S(t,r(t))=Q\cosh^{1+\alpha} t\alpha\sqrt{C} \ .
\ee
This tells that the system is unstable if there is a solution 
$0>\alpha >-1$, i.e., $0<A<{C \over 2}$. On the other hand, the 
perturbation becomes stable if there is a solution where $\alpha<-1$, 
i.e., $A>{C \over 2}$. 

The radius of the horizon $r_h$ is given by 
\be
\label{hrds}
r_h=\e^{-\phi} =\e^{-\left(\phi_0 + \epsilon S(t,r(t))\right)}\ .
\ee
Let the initial perturbation is negative $Q>0$. Then the radius shrinks 
monotonically, i.e., the black hole evaporates in case of $0>\alpha>-1$.
On the other hand, the radius increases in time and approaches to the 
extremal limit asymptotically 
\be
\label{Shrzn2}
S(t,r(t))\rightarrow Q\e^{(1+\alpha) t |\alpha|\sqrt{C}} \ .
\ee
in case of $\alpha<-1$. The latter case corresponds to the anti-evaporation
of Nariai black hole observed by Bousso and Hawking \cite{BH2}.

Eq.(\ref{det}) can be solved with respect to ${A \over C}$:
\bea
\label{Asol}
{A \over C}&=&{1 \over 2}\left(-2304 + 192 g + 36 g^2 + 24 g^2 \ln\mu^2
\right)^{-1} \nn
&& \times \left\{ -2304 + 144 g - g^2 - 6g^2 m \pm \left( 313344 g 
+ 63744 g^2 \right.\right. \nn
&& -8640 g^3 + g^4 + 27648 g \ln\mu^2 + 53568 g^2 \ln\mu^2 \nn
&& \left.\left. 
- 5556 g^3  \ln\mu^2 + 36 g^2 ( \ln\mu^2)^2 \right)^{1 \over 2}\right\} \\
&\sim&{1 \over 2} \pm {1 \over 8} g^{1 \over 2} + {\cal O}(g) 
\nn
g&\equiv&8GN\e^{2\phi_0} \nonumber
\eea 
In the classical limit $g=0$, $A={C \over 4}$, 
what tells that there does not occur any kind of the radiation 
in the solution.  Near the classical limit $g\sim 0$, there are solutions 
corresponding to both of stable and instable ones. 
It might be surprizing that there is an instable mode since the 
extremal solution is usually believed to be stable.

The global behavior of $A$ as a function of $g\equiv 8GN\e^{2\phi_0}$ and 
$\ln\mu^2$ is given in Figures. In Fig.1, the vertical line corresponds to 
${A \over C}$ and the horizontal one to $g$ when $\ln\mu^2=1$. There is a 
singularity near $g\sim 5$. In the wide range from $g=0$ to near the 
singularity, there coexist the stable and instable modes. Near the 
singlularity, both of the modes become stable ones. Beyond the singularity, 
${A \over C}$ becomes negative and there is no consistent solution. In 
Fig.2a and 2b, we show the global behavior of two modes as a function of 
$g\equiv 8GN\e^{2\phi_0}$ and $\ln\mu^2$. Fig.2a corresponds to the 
instable mode and 2b to the stable one. In Fig.2a, the singularity 
appeared in Fig.1 disappears near $g=3$.

\section{ Quantum corrections to BH parameters}

We investigate now the quantum corrections to mass, charge, etc. by 
considering the limit of $\phi\rightarrow -\infty$. In the large $N$ 
limit, the leading order is proportional to $N$. Since the leading order 
is the contribution of the one-loop correction it is also proportional to 
the effective gravitational constant $\e^{2\phi_0}$. Therefore the quantum 
correction under consideration appears in the form of $N\e^{2\phi_0}$. 
Since $\e^{-\phi_0}$ is the scale of the black hole radius in the 
corresponding 4 dimensional model, the above expansion also corresponds to 
the large radiusexpansion. Therefore we can assume that the quantum 
corrections for $\rho$ and $\phi$ appear in the following form
\be
\label{pert2}
\rho=\rho_{cl} + 8GN\e^{2\phi_0}\rho_1\ ,\ \ \ 
\phi=\phi_{cl} + 8GN\e^{2\phi_0}\phi_1
\ee
neglecting the second and higher power of $N\e^{2\phi_0}$. Here 
$\rho_{cl}$ and $\phi_{cl}$ are the classical parts of $\rho$ and $\phi$, 
respectively, which are given by (see (\ref{MNYmetric}))
\bea
\label{clmetric}
\e^{2\rho_{cl}}&=& 1-2m \e^{-2\lambda x} 
+ q^2 \e^{-4\lambda x} \nn
\e^{-2(\phi_{cl}-\phi_0)}&=&\e^{2\lambda x}\ .
\eea
Since the limit of $\phi\rightarrow -\infty$ means the large radius of 
black hole, we expand $\rho_{cl}$ and $\phi_{cl}$ as power series of 
$\e^{-2\lambda x}$. Using the Eqs.(\ref{eqnpp2}), (\ref{req2}) and 
(\ref{eqtp2}) and the boundary conditions 
\bea
\label{bc}
\rho_1,\ \phi_1\rightarrow 0\ \ &&\mbox{when}\ \ x\rightarrow -\infty \\
(\rho\rightarrow \rho_{cl}\ ,\ \ \phi\rightarrow \phi_{cl})\ , \nonumber
\eea
we find 
\bea
\label{asybh}
\rho_1&=&-\left({7 \over 32} + {1 \over 4}\ln\mu^2\right)
\e^{-2\lambda x} - \left( {5 \over 12} + {1 \over 2}\ln\mu^2 
\right)\e^{-4\lambda x} + {\cal O}(\e^{-6\lambda x}) \nn
\phi_1&=&-{1 \over 32}\e^{-2\lambda x}
+\left({13 \over 96}+{1 \over 16}\ln\mu^2\right)
m\e^{-4\lambda x}+ {\cal O}(\e^{-6\lambda x}) \nn
t_0&=&-\lambda^2GN (1 + \ln \mu^2)\ .
\eea
Here we change the radial coordinate by 
\be
\label{chcoord} 
dx =\e^{2\rho}dr\ ,
\ee
which gives the metric of the form 
\be
\label{Gmetric}
ds^2=-G(x)dt^2 + {1 \over G_(x)}dx^2\ ,\ \ G(x)\equiv\e^{2\rho(r(x))}\ .
\ee
Since the parameter $m$ related to the black hole mass is classically 
given by 
\be
\label{rhocl}
\rho_{cl}=-m\e^{-2\lambda x}+{\cal O}(\e^{-4\lambda x})\ ,
\ee
we can find the quantum correction $\delta m$ to $m$ as 
\be
\label{dm}
\delta m=\left({7 \over 32}+{1 \over 4}\ln\mu^2
\right)8GN\e^{2\phi_0}\ .
\ee
If we use the mass formula from \cite{MNY}
\be
\label{mass}
M={8 \over \sqrt{\alpha'}}m\e^{-2\phi_0}\ ,
\ee
$\delta m$ gives the quantum corection to mass,
\be
\label{dM}
\delta M={8 \over \sqrt{\alpha'}}
\left({7 \over 32}+{1 \over 4}\ln\mu^2\right)8GN\e^{2\phi_0}\ .
\ee
On the other hand, (\ref{eqA2}) and (\ref{asybh}) tell that the asymptotic 
behavior of the electric field is not changed by the quantum correction.
This would restrict parameter $q$, which is related with $B$ by
\be
\label{Bq}
B=\lambda q \e^{-2\phi_0}\ ,
\ee
and the charge of black hole would not change.

Eq.(\ref{asybh}) tells
\bea
\label{G}
G(x)&\equiv&\e^{2\rho} \nn
&=& 1+2\left\{-m - \left({7 \over 32} + {1 \over 4}\ln\mu^2\right)
8GN\e^{2\phi_0}\right\}\e^{-2\lambda x} \nn
&& + \left\{q^2 - \left( {5 \over 6} + {1 \over 2}\ln\mu^2 
\right)8GN\e^{2\phi_0}
\right\}\e^{-4\lambda x} + {\cal O}(\e^{-6\lambda x}) \ ,
\eea
what gives the shift of the position of the horizons
\bea
\label{horizonshift}
\e^{2\lambda x^\pm}&=&m\pm\sqrt{m^2 - q^2} \\
&& +8GN\e^{2\phi_0}\left\{-\left({7 \over 32}
+ {1 \over 4}\ln\mu^2\right)\mp{m \over \sqrt{m^2 - q^2}}\left(
{1 \over 96}+{1 \over 4}\ln\mu^2 \right)\right\}\ .\nonumber
\eea
When the metric has the form of (\ref{Gmetric}) and $G(x)\rightarrow 1$ 
when $x\rightarrow +\infty$, the temperature $T$ 
of the black hole is given by,
\be
\label{tempe}
T={1 \over 4\pi}\left.{\left|{d G \over dx}\right|}\right|_{x=x^+}\ .
\ee
Using (\ref{G}), we find
\bea
\label{tmpG}
4\pi T&=&{4\lambda \over q^2}\left\{-m^2+q^2+m\sqrt{m^2-q^2}\right\} \nn
&&+32\lambda GN \e^{2\phi_0}\left[-{7 \over 32}-{1 \over 4}\ln\mu^2
+ {m^2 \over q^2}\left({5 \over 6}+{1 \over 2}\ln\mu^2\right) \right. \\
&& \left. +{1 \over q^2\sqrt{m^2-q^2}}\left\{-\left({5 \over 4}
+{3 \over 2}\ln\mu^2\right)m^3
+\left({1 \over 48}+{1 \over 2}\ln\mu^2\right)q^2m\right\}\right]\ .
\nonumber
\eea

Recently Teo has found that the model of McGuigan, Nappi and Yost is 
related with five dimensional extremal black hole in Type II superstring 
theory \cite{T} by the duality. The five dimensional black hole can be 
obtained as a solution of ten dimensional Type II supergravity when five 
space coordinates $(x_5, x_7, \cdots x_9)$ are compactified. The metric has 
the following form:
\bea
\label{5dmetric}
ds^2&=& - (H_1K)^2 f dt^2 + H_1^{-1}K \left( dx_5 - ({K'}^{-1} 
-1)dt\right)^2  \nn
&& + H_5\left( f^{-1}dr^2 + r^2 d\Omega_3^2\right) +\sum_{i=6}^9 dx_i^2 \ .
\eea
Here $d\Omega_3^2$ is the metric on the unit three-sphere and 
\bea
\label{HK}
&r^2=\sum_{i=1}^4 x_i^2\ ,\ & f=1-{r_0^2 \over r^2}\ , \nn
&H_1=1 + {r_0\sinh^2\alpha \over r^2}\ ,\ \ 
&H_1'=1 - {r_0\sinh\alpha\cosh\alpha \over r^2}H_1^{-1}\ ,\nn
&H_5=1 + {r_0\sinh^2\beta \over r^2}\ ,\ \ 
&H_5'=1 - {r_0\sinh\beta\cosh\beta \over r^2}H_5^{-1}\ ,\nn
&K=1 + {r_0\sinh^2\gamma \over r^2}\ ,\ \ 
&K'=1 - {r_0\sinh\gamma\cosh\gamma \over r^2}K^{-1}\ .
\eea
The entropy is, as usually, given by
\be
\label{5entropy}
S={A_H \over 4G_5}={2\pi^2r_0^3\cosh\alpha\cosh\beta\cosh\gamma 
\over 4G_5}\ .
\ee
Here $A_H$ is the horizon area and $G_5$ is Newton constant in five 
dimensions. The extremal limit is obtained by taking $r_0\rightarrow 0$ 
but keeping $r_0\sinh\alpha$, $r_0\sinh\beta$ and $r_0\sinh\gamma$ to be 
finite. When taking the extremal limit and setting $\alpha=\gamma$, it was 
shown \cite{T} that the metric (\ref{5dmetric}) is dual to the following 
metric 
\be
\label{dualmetric}
ds^2=-\left(1+{r_1^2 \over r^2}\right)^{-2}dt^2+{r_5^2 \over r^2}dr^2 
+ r_5^2 d\Omega_3^2 + + \sum_{i=5}^9dx_i^2 \ .
\ee
Here
\be
\label{r15}
r_1=r_0\sinh\alpha\ ,\ \ \ r_2=r_0\sinh\beta\ .
\ee
The metric is direct product of 5 dimensional compact space corresponding 
to $(x_5,x_6,\cdots,x_9)$, three sphere of the radius $r_5$ and 
two-dimensional spacetime. Changing the coordinate
\be
\label{Ccoord}
\sqrt{m^2-q^2} \e^{-\lambda x}={r_0^2 \over r^2 + r_1^2}\ ,\ \ \ 
r_0={1 \over \lambda}\ ,
\ee
the metric of the two-dimensional spacetime becomes that of 
(\ref{MNYmetric}). It was also shown that the entropy $S$ in 
(\ref{5entropy}) is given in two dimensional language as follows
\be
\label{2entropy}
S=4\pi\e^{-2\phi_0}\e^{\lambda x^+}\ .
\ee
Using the quantum correction in (\ref{horizonshift}), we find the quantum 
correction to the entropy
\bea
\label{entropy}
S&=&4\pi\e^{-2\phi_0}\Bigl[m+\sqrt{m^2 - q^2} \nn
&& +8GN\e^{2\phi_0}\left\{-\left({7 \over 32}
+ {1 \over 4}\ln\mu^2\right)-{m \over \sqrt{m^2 - q^2}}\left(
{1 \over 96}+{1 \over 4}\ln\mu^2 \right)\right\}\Bigr]\ .
\eea
Hence, we demonstrated that two-dimensional considerations may be useful 
to define quantum corrections not only to two-dimensional parameters 
but also to five-dimensional ones (like entropy).


\section{Kaluza-Klein reduction of 3d gravity}

Let us apply above technique to
 BTZ BHs. Three dimensional Einstein gravity with the 
cosmological term has the exact rotating black hole solution \cite{BTZ} 
and the solution can be regarded as an exact solution of string theory 
\cite{H}. In this section, we first consider the circle reduction of three 
dimensional Einstein gravity to two dimensional one. Now we identify the 
coordinates as $x^1=t$, $x^2=r$, $x^3=\phi$. Here $t$, $r$ and $\phi$ are 
the coordinates of time, radius, and angle. We now assume that all the 
fields do not depend on $x^3$. Then, under the following infinitesimal 
coordinate transformation $x^3\rightarrow x^3 + \epsilon (x^1,x^2)$, the 
metric tensors transform as follows;
\be
\label{metrictrnsf}
\delta g^{(3)3\alpha}=\delta g^{(3)\alpha 3}=
g^{(3)\alpha\beta}\partial_\beta\epsilon\ ,\ \ 
\delta g^{(3)33}=2g^{(3)3\alpha}\partial_\alpha\epsilon\ ,\ \ 
\delta g^{(3)\alpha\beta}=0
\ee
Here $\alpha, \beta=1,2$. 
Eq.(\ref{metrictrnsf}) tells that we can identify the gauge vector field 
$A_\alpha$ with $g^{(3)3\alpha}=g^{(3)\alpha\beta}A_\beta$. 
Eq.(\ref{metrictrnsf}) also tells that the metric tensor $g^{(3)\mu\nu}$ 
($\mu,\nu=1,2,3$) can be parametrized a la Kaluza-Klein as
\be
\label{3dmtrc}
g^{(3)\mu\nu}=\left(\begin{array}{cc}
g^{\alpha\beta} & g^{\alpha\gamma}A_\gamma \\
g^{\beta\gamma}A_\gamma & \e^{2\phi} +g^{\gamma\delta}A_\gamma A_\delta \\
\end{array}\right)\ .
\ee
Then we find 
\bea
\label{g3}
g^{(3)}&\equiv&{1 \over \det g^{(3)\mu\nu}}=g\e^{-2\phi} \ \ 
(g\equiv {1 \over \det g^{\mu\nu}}) \\
\label{ginv}
g^{(3)}_{\mu\nu}&=&\left(\begin{array}{cc}
g_{\alpha\beta}+\e^{-2\phi}A_\mu A_\nu & -A_\gamma \\
-A_\gamma & \e^{-2\phi} \\
\end{array}\right)\ .
\eea
In the following, the quantities in three dimensions are denoted by the 
suffix ``$(3)$'' and the quantities without the suffix are those in two 
dimensions unless we mention. In the parametrization (\ref{3dmtrc}),  
curvature has the following form 
\be
\label{cvsclr}
R^{(3)}=R+\Box \phi - \partial_\alpha\phi \partial^\alpha\phi - {1 \over 4}
\e^{-2\phi}F_{\alpha\beta}F^{\alpha\beta}\ .
\ee
Here $F_{\alpha\beta}$ is the field strength: $F_{\alpha\beta}
=\partial_\alpha A_\beta - \partial_\beta A_\alpha$. Then the action $S$ 
of the gravity with cosmological term and with  $N$ free scalars $f_i$ 
($i=1,\cdots,N$) in three dimensions is reduced as
\bea
\label{redS}
S&=&{1 \over 4\pi^2} 
\int d^3x \sqrt{g^{(3)}}\left\{{1 \over 8G}\left( R^{(3)} + \Lambda \right)
+ {1 \over 2}\sum_{i=1}^N \partial_\mu f_i \partial^\mu f^i \right\}\nn
&\sim&{1 \over 2\pi} \int d^2x \sqrt{g}\e^{-\phi}\left\{
{1 \over 8G}\left(R - {1 \over 4} \e^{-2\phi}F_{\alpha\beta}F^{\alpha\beta}
+\Lambda \right)+ {1 \over 2}\sum_{i=1}^N \partial_\alpha f_i 
\partial^\alpha f^i \right\}\ .
\eea
Note that the contribution from the second and third terms in 
(\ref{cvsclr}) becomes the total derivative which maybe neglected.
One BH solution for the classical action is given by \cite{BTZ}
\bea
\label{BTZclass}
&& ds^2=-\e^{2\rho_{cl}}dt^2 + \e^{-2\rho_{cl}}dx^2 \ ,\ \ 
\e^{2\rho_{cl}}=-M+{x^2 \over l^2} + {J^2 \over 4x^2} \nn
&& \e^{-\phi}=\e^{-\phi_{cl}}=\e^{-\phi_0}x
\eea
Here ${1 \over l^2}={\Lambda \over 2}$. $M$ is the parameter related to 
the black hole mass and $J$ is that of the angular momentum in 3D model or 
the electromagnetic charge in the 2D model. The extremal limit is given by 
$J^2=l^2 M^2$. In order to consider this limit, we change the coordinates 
as follows
\be
\label{chgcoord}
x^2={l^2 M \left( 1 + \epsilon \tanh r \right) \over 2}\ ,\ \ \ 
t={l \over \epsilon\sqrt{2}}\tau\ .
\ee
Then $r\rightarrow +\infty$ corresponds to outer horizon and 
$r\rightarrow -\infty$ corresponds to inner horizon.  Taking the limit of 
$\epsilon\rightarrow 0$, we obtain
\be
\label{extsol}
ds^2={l^2 M \over 4\cosh^2 r}\left(d\tau^2 - dr^2\right)\ ,\ \ \ 
\e^{-\phi}={l\e^{-\phi_0}}{\sqrt{M \over 2}}\ .
\ee
Note that $\phi$ becomes a constant in this limit.

Let us now discuss the quantum corrections induced by $N$ free conformally 
invariant dilaton coupled scalars $f_i$:\footnote{Quantum corrections due 
to matter near BTZ black hole have been also studied in \cite{BVZ}.}. 
We use (\ref{qc}) and (\ref{Gamma1b}) by replacing $2\phi\rightarrow \phi$.
In the study of quantum corrected BH we work in the conformal gauge 
(\ref{cg}). We are going to study (in)stability of black holes in the 
same way as in four dimensions \cite{NO2}.

The equations of motion with account of quantum corrections are given by
\bea
\label{eqnppBTZ}
0&=&{1 \over 8G}\e^{-\phi}\left(-\partial_\pm \rho \partial_\pm\phi 
- {1 \over 2} \partial_\pm^2\phi + {1 \over 2} (\partial_\pm\phi)^2\right) 
+{N \over 12}\left( \partial_\pm^2 \rho 
- \partial_\pm\rho \partial_\pm\rho \right) \nn
&& +{N \over 8} \left\{\left( \partial_\pm \phi \partial_\pm\phi \right)
\rho+{1 \over 2}{\partial_\pm \over \partial_\mp}
\left( \partial_\pm\phi \partial_\mp\phi \right)\right\} \nn
&& +{N \over 8}\left\{ 2 \partial_\pm \rho \partial_\pm \phi 
-\partial_\pm^2 \phi \right\} + t^\pm(x^\pm) 
+{N \over 8} \left( \partial_\pm \phi\partial_\pm\phi \right)\ln \mu^2 \\
\label{reqBTZ}
0&=&{1 \over 8G}\e^{-\phi}\left(4\partial_+ \partial_- \phi 
- 4 \partial_+\phi\partial_-\phi 
- \Lambda \e^{2\rho} + 2\e^{-2\phi-2\rho}(F_{+-})^2\right) \nn
&& -{N \over 12}\partial_+\partial_- \rho
-{N \over 16}\partial_+ \phi \partial_- \phi 
+{N \over 8}\partial_+\partial_-\phi \\
\label{eqtpBTZ}
0&=& {1 \over 8G}\e^{-\phi}\left(4\partial_+ \partial_- \rho
+ {\Lambda \over 2} \e^{2\rho}
+ 3\e^{-2\phi-2\rho}(F_{+-})^2\right)  \nn
&& -{N \over 16}\partial_+(\rho \partial_-\phi)
-{N \over 16}\partial_-(\rho \partial_+\phi)
-{N \over 8}\partial_+\partial_-\rho  
-{N \over 4} \partial_+ \partial_- \phi \ln \mu^2\\
\label{eqABTZ}
0&=&\partial_\pm(\e^{-3\phi-2\rho}F_{+-})\ .
\eea
Eq.(\ref{eqABTZ}) can be integrated to give
\be
\label{eqA2BTZ}
\e^{-3\phi-2\rho}F_{+-}=B\ (\mbox{constant})\ .
\ee
We now assume $\phi$ is constant as in the classical extremal solution 
(\ref{extsol}) even when we include the quantum correction:
$\phi=\phi_0\ (\mbox{constant})$. Then we obtain
\be
\label{constRBTZ}
B^2={\Lambda \e^{-4\phi_0} \over 2 } \ ,\ \ 
R=R_0
\equiv -{4\Lambda^2 \e^{-\phi_0} \over 
\e^{-\phi_0}-{GN \over 4} }\ .
\ee
Quantum corrected extremal (static) solution corresponding to the 
classical one (\ref{extsol}) can be written as in (\ref{rho0}) 
($r={x^+ - x^- \over 2}$).

In order to investigate the (in)stability in the above solution, we 
consider the perturbation around the extremal solution as in (\ref{pert}).
We neglect the second term in (\ref{qc}) again. Then (\ref{reqBTZ}) and 
(\ref{eqtpBTZ}) (and (\ref{eqA2BTZ})) lead to the following equations:
\bea
\label{reqpertBTZ}
0&=&{1 \over 8G}\e^{-\phi_0}\left\{4\partial_+\partial_-S 
-\Lambda \e^{2\rho_0} (2R-S) + 2B^2 \e^{2\rho_0+4\phi_0}(2R+3S)\right\} \nn
&& -{N \over 12}\partial_+\partial_- R+ {N \over 8}\partial_+\partial_-S \\
\label{eqtppertBTZ}
0&=&{1 \over 8G}\e^{-\phi_0}\left\{4\partial_+\partial_-R
- 4\partial_+\partial_- \rho_0 S
+ {\Lambda \over 2} \e^{2\rho_0} (2R-S)\right.\nn
&& \left. + 3B^2 \e^{2\rho_0+4\phi_0}
(2R+3S)\right\} - {N \over 8}\partial_+\partial_- R 
-{N \over 4}\ln \mu^2 \partial_+\partial_- S \ .
\eea
By assuming $R$ and $S$ have the form of (\ref{RS}), we can rewrite 
(\ref{reqpertBTZ}) and (\ref{eqtppertBTZ}) as algebraic equations and the 
condition that there is  non-trivial solution for $Q$ and $P$ is given by
\bea
\label{detBTZ}
F^{BTZ}(A)&\equiv&\e^{-2\phi_0}\left({2R_0 \over C}A + 4\Lambda\right)^2 
- \left({GNR_0 \over C}A\right)^2 \nn
&& + {GNR_0 \over 3C}A\left\{\e^{-\phi_0}\left(-{R_0 \over 2}
+ 4\Lambda\right) - GN (\ln\mu^2) {R_0 \over C}A\right\}=0\ .
\eea 
Here $A={\alpha(\alpha -1 ) C \over 4}$. $\e^{-\phi}$ is identified with 
the radius in 3D model, so horizon is given by the condition of 
(\ref{horizon}). 

Eq.(\ref{detBTZ}) can be solved with respect to $A$,
\bea
\label{AC}
{A \over C}&=&\left[ -4096 - 2176 g + 48 g^2 
\pm \left\{ 18939904 g + 4321280 g^2 \right.\right.\nn
&& - 208832 g^3 + 2304 g^4 + \left(786432 g - 49152 g^2 \right. \nn
&& \left.\left.\left. + 768 g^3 \right)\ln\mu^2
\right\}^{1 \over 2}\right] 
 \times \left\{2\left( -4096 + 16 g^2 + 192 g^2 m \right)\right\}^{-1}\nn
&\sim&{1 \over 2} \pm {3 \sqrt{2} \over 4}g^{1 \over 2} 
+ {\cal O}(g) 
\eea
Here $g \equiv 8GN\e^{\phi_0}$. Note that there are two solutions near the 
classical limit $g\rightarrow 0$, which correspond to stable and instable 
modes, respectively. It might be surprizing that there is an instable mode 
since the extremal solution is usually believed to be stable. The global 
behavior of ${A \over C}$ is given in Figures. In Fig.3, the vertical line 
corresponds to ${A \over C}$ and the horizontal one to $g$ when 
$\ln\mu^2=1$. There is a singularity when $g\sim 0.4376$. The singularity 
occurs when the denominator in (\ref{AC}) vanishes. In the range from 
$g=0$ to near the singularity, there coexist the stable and instable 
modes. In Fig.4a and 4b, we present the global behavior of two modes as a 
function of $g\equiv 8GN\e^{g\phi_0}$ and $\ln\mu^2$. Fig.4a corresponds 
to the stable mode and 2b to the instable one.


\section{Higher order perturbations and multiple BHs}

Recently Bousso\cite{B} has shown the possibility that de Sitter space 
disintegrates into an infinite number of copies of de Sitter space. 
In the scenario, Schwarzschild de Sitter black hole becomes an extremal 
one which is known to be the Nariai solution. The topology of the Nariai 
space is $S^1\times S^2$, therefore the solution can be regarded to 
express the topology of a handle. If there is a perturbation, multiple 
pair of the cosmological and black hole horizons are formed. After that 
the black holes evaporate and the black hole horizons shrink to vanish.
Therefore the handle is separated into several pieces, which are copies 
of the de Sitter spaces.

The metric (\ref{exmetric}) or (\ref{extsol}) in the extremal limit or 
its quantum analogue (\ref{rho0}) has the same spacetime structure as the 
Nariai solution except its signature. Therefore there is a possibility of 
the multiple production of universes from the extremal soltion. 

If we define new coordinate $\theta$ by
\be
\label{theta}
\sin\theta = \tanh \left(r\sqrt{C}\right)
\ee
the metric corresponding to (\ref{rho0}) becomes
\be
\label{thetametric}
ds^2={2C \over R_0}\left(-dt^2 + {1 \over C}d\theta^2\right)\ .
\ee
Since $\theta$ has the period $2\pi$, the topology of the space with 
the metric (\ref{thetametric}) can be $S_1$. Note that Eq.(\ref{theta}) 
tells there is one to two correspondence between $\theta$ and $r$.

The operator 
\be
\label{Delta}
\Delta\equiv C\cosh^2\left(r\sqrt{C}\right)\partial_+\partial_- 
\ee
in Eq.(\ref{eigenRS}) can be regarded as the Laplacian on two-dimensional 
Lorentzian hyperboloid, i.e., the Casimir operator of $SL(2,R)$. The 
raizing and lowering operators $L_\pm$ of $SL(2,R)$ are given by
\be
\label{rl}
L_\pm=\e^{\pm t\sqrt{C}}\left(\sinh\left(r\sqrt{C}\right)
{\partial \over \partial t} \pm \cosh\left(r\sqrt{C}\right)
{\partial \over \partial r}\right)\ .
\ee
The eigenfunction of $\Delta$ in $S$ and $R$ (\ref{RS}) is the sum of 
highest and lowest weight representation 
$\e^{\pm t\alpha\sqrt{C}}\cosh^\alpha r\sqrt{C}$. 
We now consider the following eigenfunction
\bea
\label{S2}
S(t,r)&=&{Q \over 4} \left(L_+\e^{ t\alpha\sqrt{C}}\cosh^\alpha r\sqrt{C}
+ L_-\e^{- t\alpha\sqrt{C}}\cosh^\alpha r\sqrt{C}\right) \nn
&=&Q\sinh\left(t(\alpha+1)\sqrt{C}\right)\cosh^\alpha r\sqrt{C}
\sinh r\sqrt{C}\ .
\eea
Since the evaporation of the black hole in \cite{B} corresponds to the 
instable mode, we restrict $\alpha$ to be $0>\alpha>-1$. Then the 
condition of horizon (\ref{horizon}) gives
\be
\label{horizonS2}
\tanh\left(r\sqrt{C}\right)=\pm{\alpha+1 - 
\sqrt{(\alpha + 1)^2 - 4\alpha\tanh^2\left(t(\alpha+1)\sqrt{C}\right)} 
\over 2\alpha\tanh\left(t(\alpha+1)\sqrt{C}\right)}\ .
\ee
We find the asymptotic behaviors of $S$ are given by
\bea
\label{Sasym}
S(t,r(t))&\stackrel{t\rightarrow 0}{\rightarrow}&\pm (\alpha +1)CQ 
t^2 \nn
&\stackrel{t\rightarrow +\infty}{\rightarrow}&\pm{Q \over 2^{\alpha+2}}
\left({1-\alpha \over 1+\alpha }\right)^{\alpha + 1 \over 2}
\e^{t(\alpha + 1)(\alpha + 2)\sqrt{C}}
\eea
Since the radius of the horizon is given by (\ref{hrds}), the horizon 
corresponding to $+$ sign in (\ref{Sasym}) grows up to infinity and that 
to $-$ sign shrinks to vanish when $t\rightarrow +\infty$ although the 
perturbation does not work when $S$ is large. The horizon of $+$ sign 
would corespond to outer horizon and that of minus sign to inner horizon. 
Since there is one to two correspondence between the radial coordinates 
$r$ and $\theta$, there are two outer horizons and two inner ones. When we 
regard the model as the model reduced from four (three) dimensions, the 
handle with the topology of $S_1\times S_2$ ($S_1\times S_1$) is separated 
to two pieces when the two inner horizons shrink to vanish. This result 
would be generalized if we use higher order eigenfunctions as perturbation
\be
\label{Sn}
S(t,r)={Q \over 4}\left((L_+)^n \e^{ t\alpha\sqrt{C}}\cosh^\alpha r\sqrt{C}
+ (L_-)^n\e^{- t\alpha\sqrt{C}}\cosh^\alpha r\sqrt{C}\right)\ .
\ee
 From the coorespondence with Bousso's work \cite{B}, there would appear 
$2n$ outer horizons and $2n$ inner horizons and the spacetime could be 
disintegrated into $2n$ pieces when the inner horizons shrink to vanish. 
Of course, the confirmation of this effect should be checked with some 
other methods. It is also interesting to note that proliferation of de 
Sitter space is somehow different from baby universes creation based also 
on the topological changes (for a review,see ref.\cite{GWG}) as it occurs 
at large scales \cite{B}.
Note that horizon equation for the higher mode is $2n$-th order 
polynomial with respect to $\tanh r\sqrt{C}$, which is the 
reason why it is expected that there can be $2\times 2n$ 
horizons. However, it is difficult to confirm that all solutions 
are real solutions.

\section{Discussion}
We studied quantum properties of 2D charged BHs and BTZ BH. The quantum 
effects of dilaton coupled scalars in large $N$ approximation have been 
taken into account. Quantum corrected solution for 2D charged BH has been 
found and quantum corrections to mass, charge, Hawking temperature and 
BH entropy have been evaluated. As 2D dilaton-Maxwell gravity with 
non-minimal scalars under discussion may be also considered as toy (or 
spherically reduced) version for 4D or 5D Einstein-Maxwell-minimal scalar 
theory then higher dimensional interpretation of results may be done. (It 
is also possible due to duality of 2D BH with 5D BH). Quantum 
(in)stabilities of 2D charged BH and BTZ BH are discussed and it is shown 
the presence in the spectrum of the mode corresponding to anti-evaporation 
like in the case of 4D SdS BH. Hence, the important qualitative result 
of our work is that we show:quantum anti-evaporation of BHs 
(even BTZ BHs) is quite 
general effect which should be expected for different multiple BHs (and 
not only for SdS BH). Numerical calculation shows that evaporation is 
``more" stable effect, nevertheless. The interpretation of higher modes 
perturbations as the reason of disintegration of the space itself to 
copies is given. It is clear that in order to understand if the above 
effects may really happen in early universe one should select boundary 
conditions corresponding to BH formation and analyse above processes 
subject to such boundary conditions. Depending on the choice of boundary 
conditions they may be compatible with anti-evaporation as it happens for 
Nariai anti-evaporating BH \cite{NO2}.

It is not difficult to generalize the results of this work to other higher 
dimensional BHs where metric may be presented as the product of 2D charged 
BH with D-2-dimensional sphere. Then large $N$ approximation and spherical 
reduction may be applied as well 
(as it was done in 3D case) and there are no principal problems to 
study quantum properties of such BHs (at least in large $N$ approximation). 
>From another point, duality may help in the study of connections between 
low and higher dimensional BHs on quantum level. Number of new effects are 
expected to be found in such studies.

It could be very interesting also to consider supersymmetric 
generalization of above results. It seems to be possible to do as anomaly 
induced effective action for dilaton coupled supersymmetric 2D or 4D 
theories is known \cite{NOG}.

\noindent
{\bf Acknoweledgments.} We thank R. Bousso and S.Hawking
 for some discussion.
The work by SDO has been partially supported by RFBR, Project No.96-02-
16017.

\

\newpage

\noindent
{\Large\bf Figure Captions}

\ 

\noindent
Fig.1 ${A \over C}$ (vertical line) versus 
$g\equiv 8GN\e^{2\phi_0}$ (horizontal line)
when $\ln\mu^2=1$.

\ 

\noindent
Fig.2a,b Two branches of ${A \over C}$ (vertical line) versus 
$g\equiv 8GN\e^{2\phi_0}$ in $[0,10]$ and $\ln\mu^2$ in
$[0,3]$. Fig.2a corresponds to the instable mode and 2b to the 
stable one.


\ 

\noindent
Fig.3 ${A \over C}$ (vertical line) versus 
$g\equiv 8GN\e^{\phi_0}$ (horizontal line)
when $\ln\mu^2=1$.

\ 

\noindent
Fig.4a,b Two branches of ${A \over C}$ (vertical line) versus 
$g\equiv 8GN\e^{\phi_0}$ in $[0,1]$ and $\ln\mu^2$ in
$[0,3]$. Fig.4a corresponds to the stable mode and 4b to the 
instable one.

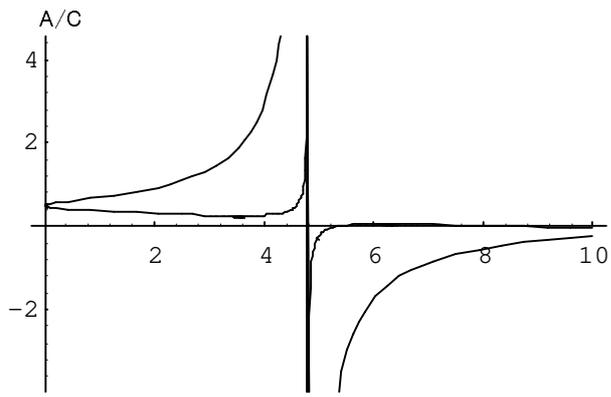
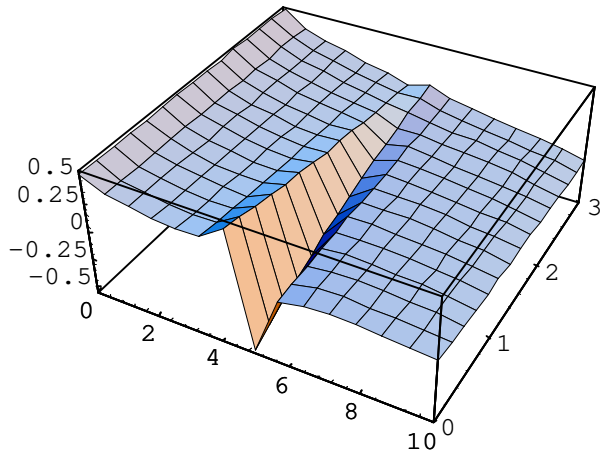
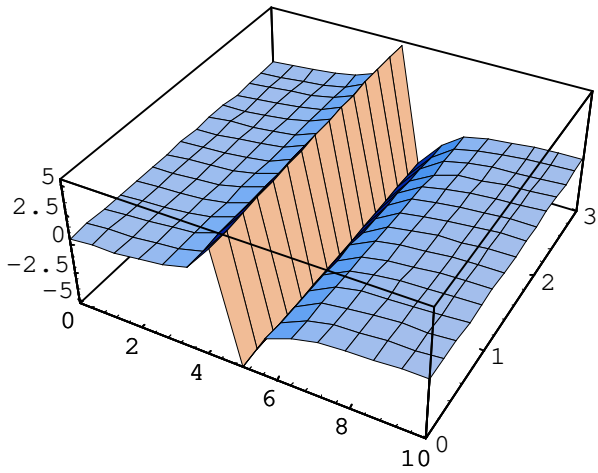
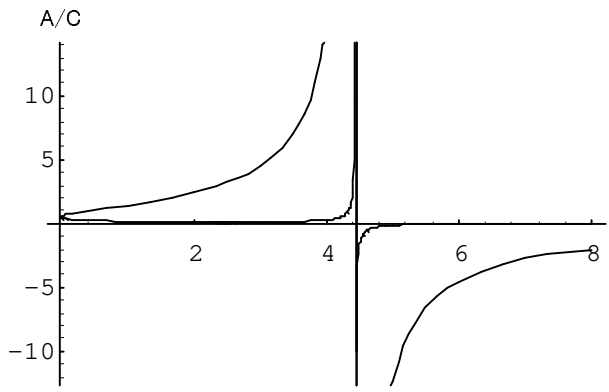
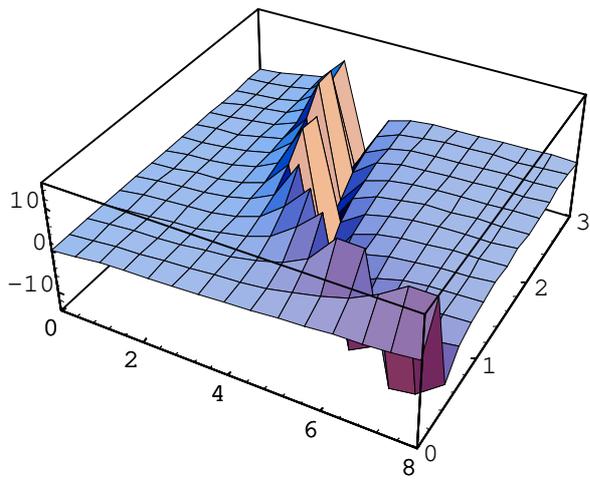
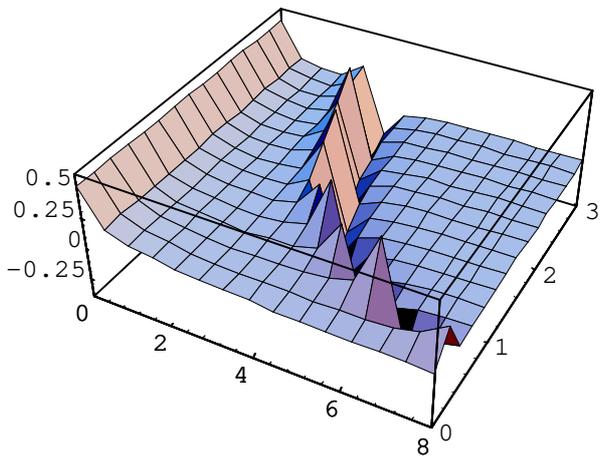
\end{document}